# JADE – A PLATFORM FOR RESEARCH ON COOPERATION OF PHYSICAL AND VIRTUAL AGENTS

Wiktor B. Daszczuk, Jerzy Mieścicki
Institute of Computer Science, Warsaw University of Technology
Nowowiejska Str. 15/19, Warsaw 00-665, Poland,
E-mail: wbd@ii.pw.edu.pl

**ABSTRACT**

In the ICS, WUT a platform for simulation of cooperation of physical and virtual mobile agents is under development. The paper describes the motivation of the research, an organization of the platform, a model of agent, and the principles of design of the platform. Several experimental simulations are briefly described.

**INTRODUCTION**

In the Institute of Computer Science, WUT the research on systems of mobile agents was initiated few years ago. The general goal of the project was (and still is) to create and to investigate models of 'social behavior' in groups of simple but physically existing robots, able to move and act in a physical environment. The target version of the system will finally include a number of such hardware units busily moving around.

Conceptually, such a system of interacting units can be viewed, of course, as a *multiagent system*. This provides the project the well-defined conceptual framework and places it in one of most interesting streams of research in contemporary computer and information science. However, in our approach agents are not merely abstract entities executing their *beliefs, desires and intentions* in a virtual environment. Instead, the 'world' where multiple agents manifest their activity is a two-dimensional, physical 'playground', where several types of objects can be placed: obstacles or forbidden areas to be omitted, things to be sought for, elements to be moved or carried etc. The basic aspects of agent's behavior (sensing and affecting the world, movement, communication etc.) are then generally interpreted in terms of physical phenomena, such as touch, sensitivity to light or sound, physical displacement, distance, speed vector, infrared or radio communication etc.

In the project emphasis is put on 'social' aspects of interactions among agents, e.g. competition, cooperation toward a common goal, group hierarchy and heterogeneity of roles vs. homogeneity and 'flat' organization etc.

Therefore, in contrast to the research in the field of application-oriented (e.g. industrial or space-mission) robotics, individual robots do not have to meet any challenging requirements as to their sophisticated patterns of movement, extraordinary precision or performance. They make just visible, 'hard' parts (or 'bodies') of agents, while agent's main 'intelligence' or 'mental' activity (e.g. decision making, learning, knowledge acquisition, communication etc.) is executed as a set of software processes which are not necessarily attributed to the mobile platform. Nevertheless, the design and physical implementation of a flock of inexpensive but operable mobile robots is among the cornerstones of the project.

This approach was recognized to be well tailored to the mission of the Institute of Computer Science, which involves both scientific research and teaching students in an academic level. First, the multiagent systems are interesting field of study as themselves. The problems open for research include not only issues of the implementation of communicating distributed modules but also knowledge acquisition and decision-making strategies based on learning and other DAI (Distributed Artificial Intelligence) techniques (Stone and Veloso 1997). Challenging group of problems are these related to means of communication, multi-level protocols, languages of communication on application level (ACL – Agent Communication Languages like KQML - Finn and Weber 1994 or KIF - Genesereth and Fikes 1992), including the issues of ontology of a language of communication (specification of concepts, objects, relations and boundaries in semantic model - Karp at al. 1999). The research would contribute to *theories of inter-agent interaction and sociality including issues of sociability, benevolence, preference, power, trust, teaming, norms, roles, teamwork*, etc. In addition, the following arguments should also be emphasized:

- From a viewpoint of information technology such a system makes concise and easily observable but challenging model of a *distributed system*, with all its problems, e.g. concurrency, uncertainty, communication delays, distributed algorithms, hierarchy of communication protocols etc. Therefore, it can be also an excellent illustration and test field for numerous labs and projects from the computer science curriculum,

- From the software engineering perspective, the system can be also used in research projects on the methodology of *concurrent and distributed software development*, e.g. in relation to object-oriented modeling and design, component-based technologies, Java programming techniques, CASE tools etc.
- Design and implementation of hardware part of an agent provides an opportunity to gain the experience on *embedded systems*, *low-level device programming*, *physical communication media* etc. Moreover, it allows the researchers and students to get in touch with real-world problems which often are underestimated even by computer professionals, e.g. uncertainty of sensors or effectors, power consumption, inertia, friction, control delay etc.

Last but not least, experiments with mini-societies of real-world robots are just *a fun*. This stimulates creativity and provides the researchers as well as students with an extraordinary motivation and satisfaction.

**PRELIMINARY PROJECT PHASE**

Many multiagent systems have been developed yet. Libraries and software tools are designed to support construction of multiagent systems. Many tools are publicly available, like Jackall (Cost and Finin 1998) or JATLite (http://java.stanford.edu/JATLiteDownload.htm). They deal mainly with communication between distributed agents and knowledge exchange. JAF (Horling 1998) offers facilities for simulation of agents in physical environment. It allows a user to add various modules to the model of agent. Much work was done under FIPA (foundation for Intelligent Physical Agents – http://www.fipa.org/), where some standards of were established. Many multiagent systems are announced to be compatible with FIPA standards.

However, the general idea of our project does not consist in the efficient implementation of the *specific* multiagent system (e.g. for the game of soccer) in order to compare it to other, competitive implementations. Instead, the immediate goal was to create the conceptual and technological framework allowing firstly for the *development* or *the definition of a multiagent system* (with the properties if its environment, agents' behavioral rules, tasks to be done, performance criteria etc.) which could be then pushed alive, observed, measured, modified and otherwise investigated. This (as well as the tutorial values of independent design) made us to work towards an own, original solution of the multiagent platform.

Initially, two tasks were performed in parallel. First task involved the construction of prototype of a mobile mini-robot, which could later act as a physical representation of an agent. An assumption was made that physical agents can interact with a number of similar, but virtual, simulated ones. A second task consisted in building of a prototype software platform on which the whole system may be visualized, real agents may interact with virtual ones, and through which the agents may communicate.

First prototype robot, a simple cart, was built from LEGO toy building blocks. The robot used very imprecise electrical engine, and a computer mouse was used as a sensor of a movement of the robot. The result was far from satisfactory: the robot could not even move there and back on the same path. The second attempt was more successful. It was a more sophisticated robot called *Roundie* because of its shape (Modzelewski 1998). The 'undercarriage' of the robot was built again from LEGO building blocks, this time fixed with some glue. A single board computer was installed on the robot. Output devices were two active wheels powered by small electrical engines, separately controlled; velocity and direction of revolution could be defined individually for each wheel. Two passive wheels just give support for the robot. Sensors of the Roundie were:
- infrared sensor of brightness of the floor,
- two infrared sensors of rotary velocity of active wheels, one per wheel,
- two infrared sensors directed to a front of Roundie, they inform about obstacles,
- four touch sensors (microswitches with 'whiskers'), two directed forward and other two - backward.

The experience from construction of Roundie is rather positive. The possibilities of control are enough to perform complicated tasks. The robot can even write its name on a sheet of paper using a pencil, or move little objects quite precisely. Yet the size of the robot (about 40 cm in diameter) and its cost (about one thousand Euro) are too large to construct a group of robots. The disadvantage of Roundie was also the means on communication – a twisted pair and RS-232 interface was used.

First prototype software platform supporting the simulation of mobile agents and observation of their behavior was also implemented (Zawistowski 1996), but its functionality was limited. All these preliminary designs provided us with the better understanding of needs, problems and pitfalls in the design of a multiagent platform. This allowed for the formulation of requirements for a new project, called JADE (Java Agent Development Environment - Pabiś and Zabrowski 2001), which is the main subject of the present paper.

The main conceptual elements of JADE are:
- model of a virtual agent and its interactions with the environment as well as with other agents,
- model of agents' 'world' (or the Environment Map), including the representation of virtual and physical agents in it.

The JADE itself supports:
- implementation of both models (of an agent and of the Environment Map),
- representation of physical agents in a map,
- visualization and simulation of the whole system,
- monitoring of physical and virtual agents' behavior,
- loading an executable code of virtual agent onto physical agent,
- development of the agent's behavioral model,
- monitoring the performance of a group of agents.

Simultaneously, a new prototype of an inexpensive and small, but fully operable mobile robot is under development. It will be based on Motorola 68k family processor and radio remote control unit, mounted on a light four-wheeled carriage. Java will be used for the programming of robot functions. However, as physical agents are not ready yet, the functional features as well as experiments reported below were tested using virtual agents only.

**MODEL OF AN AGENT AND ITS INTERACTIONS WITH THE ENVIRONMENT**

**Terminological remark**

In a literature, many definitions of *agents* and *multiagent systems* are used. An attempt to integrate different definitions was made in (Franklin and Graesser 1996). For the purposes of our work, the following definitions are sufficient:

- A **multiagent system** *is a loosely connected team of agents having strict rules of cooperation or competition.*
- An **agent** *is a computer module placed in a multiagent environment. It experiences the environment by means of sensors and influences the environment by means of effectors (devices). Agent is mobile: it can change its position in environment and initiate action compatible to 'physics' of a given environment. Behavior of an agent is fully autonomous. It serves to achieve agent's goals. Agents exists continuously, it can learn and communicate with other agents.*

The following features, commonly attributed to agents, characterize also agents in the JADE project:

| Autonomy | Agent is independent, it controls itself |
|---|---|
| Reactivity | Agent responds to signals from environment |
| Pro-activity | Agent initiates actions |
| Social behavior | There are interactions among agents |
| Mobility | Agent can change its position in an environment |
| Intelligence | Actions are intentional; Agents learn and optimize their behavior |
| Communication | Agent passes information, questions, answers, orders to other agents |
| Limited confidence | Incoming information may be unsure, incomplete or even distorted, thus agents may possess incomplete information with various degree of reliability |

**Model of an agent**

A model of an agent (Fig. 1) can be divided into layers. The lowest layer (consisting of three peer components) makes the *physical part* of an agent while the three upper layers (and a common memory) make its *abstract part*.

The physical part is responsible for operation of sensors, effectors and communicators. Type, number and characteristics of devices can be defined for specific types of agents at the stage of the planning of the experiment. Devices are controlled (e.g. in order to read state of sensors and execute actions of effectors) and monitored (to identify malfunctions of devices). After a malfunction is observed, the abstract part may change the scenario to achieve the goal in a different way or to inform other agents about the situation.

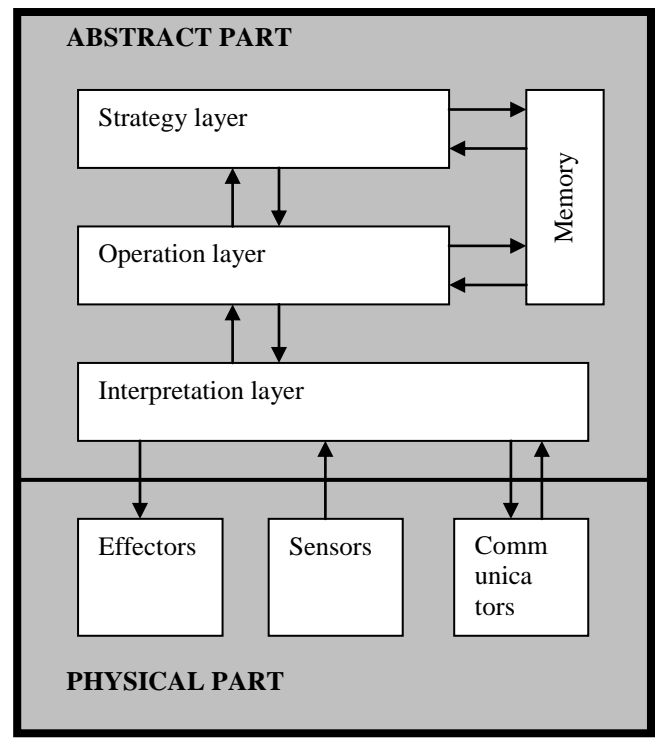

Figure 1: A Model of an Agent

The abstract part is responsible for agent's 'intelligence'. The structure of an abstract part supports effective realization of both long-term tasks (strategy) and fast responses for signals incoming from the environment. In particular:

- *Interpretation layer* communicates immediately with the physical part. It collects signals in non-processed form, interprets them as events (from predefined scope) and submits them to the operation layer. Also, interpretation layer gets orders from operation layer and converts them into sequences of actions of effectors. In this layer, a form of 'unconditioned reflexes' can be also implemented (i.e. actions performed as direct, fast responses to signals from environment, without engaging of upper abstract levels, which represent 'aware thinking').

- *Operation layer* is responsible for current decisions taking into account events from all sensors, information from other agents and several past steps of operation. The decisions are realized as sequences of orders to effectors. If other agents have to be informed, operation layer prepares messages to them and issues an appropriate command to interpretation layer.

- *Strategy layer* implements a scenario of agent behavior. The scenario is an algorithm that specifies how to achieve a goal common to a team of agents. The strategy layer uses data stored in memory: known state of the environment, known positions and current actions of other agents, the history of behavior of the agent. The format of data stored in memory depends on the type of scenario and a size of memory.

**Interactions**

Agent is an active element of environment. There are three types of agent's activity:
- Actions – influence on passive elements of environment (taking a resource, pushing an obstacle, etc.),
- Interactions – mutual influence among agents (fighting, passing resources, disturbing etc),
- Communication – passing messages to other agents.

**Environment Map**

The following elements of environment are implemented in the map:
- Obstacles (real but visualized or simulated ones), which limit freedom of movement of agents,
- Resources or passive elements, things to be sought or collected. They may be subject of competition or trade between agents,
- Agents – the only active elements, they see each other as 'mobile elements of environment'.

**Flexibility of a model**

A model of agent described above is designed to fit requirements of physical agent acting in physical environment, and in the same time to be universal enough to be compatible with virtual environment and cooperate with virtual agents. The agent may be equipped with various sets of devices and interact with other agents in multiple ways.

**THE IMPLEMENTATION OF JADE ENVIRONMENT**

**General structure**

JADE environment was designed and implemented in a client-server architecture (Fig. 2). A server (Environment Map) is an environment platform and agents are clients. Additional elements of JADE are map editor and configuration editor.

Physical and abstract parts of model of an agent are reflected in the architecture of implementation. In virtual agent all devices are simulated by the platform. After the virtual agent was tested, it may be implemented physically: all virtual devices are then replaced by physical ones and the platform stops simulating the physical environment for the agent. However, some simulation may be still necessary even for physical agent, if it has to cooperate with virtual agents and experience the result of their behavior.

Creation of an agent is very simple. Inheritance and aggregation of ready-to-use device components are typically applied. The constructor of an agent class places it directly on the map. The initial part of agent program defines devices and their initial parameters. Next, the algorithm of agent's behavior follows. Typically, this algorithm is specified in a form of a Concurrent State Machine (Mieścicki 1992). This form of specification (generally similar to UML's state diagrams, see example in Fig. 5) allows for a formal verification of many aspects of communication and cooperation among concurrent components. Rules of automatic code generation from source CSM automaton are known (Daszczuk 1998a and 1998b), and a software generator is under development.

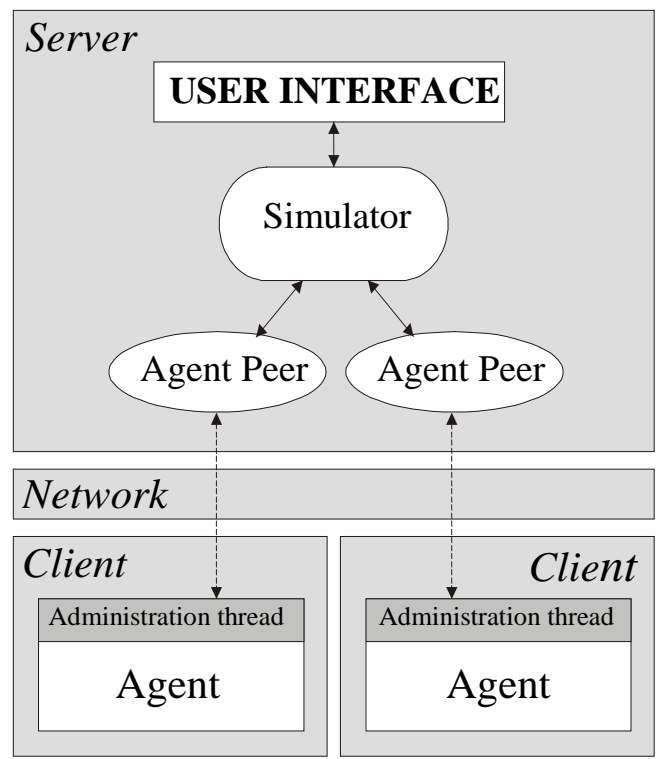

Figure 2: A Scheme of the Simulator
(Map Environment)

User can also prepare definition of 'world' in which agents will act, and configure the environment. These configuration files are stored in XML format (Bray at al. 1998) and therefore may be edited using XML edition programs.

Additional tools are designed for:
- Storing and retrieving logs of agents' behavior,
- Monitoring of internal state and sequences of actions in individual agents.

Elements of JADE platform can act either in *on-line* or *off-line mode*. In an on-line mode, an Environment Map acts as simulator for virtual agents and as visualization tool for physical agents. It serves agents and displays changes in environment. In an off-line mode, Environment Map does not respond to signals from agents and interacts with map editor, configuration editor or log player. Off-line tools may be invoked during the simulation, but the changes take effect after finishing of the simulation.

**The Environment Map– simulator**

There is an obvious tradeoff between a performance of simulator and a detailed 'reenactment' of the features of real world. Therefore, all features of real world that are not important in research on social aspects of a multiagent system are rejected or simplified. The world is two-dimensional, as the third dimension significantly complicates the model while a two-dimensional world is sufficient enough from the viewpoint of the research on agents' cooperation mechanisms, artificial intelligence algorithms etc.

The simulator is responsible for:
- simulation of physical environment:
    o change of agents' positions (in memory of the simulator and in a window on the screen),
    o detection of collisions of agents with obstacles or other agents,
    o visualization of other objects;
- communication:
    o passing messages between agents,
    o identification of agents by names,
    o arranging groups of agents.

Among the limitations of the simulator are the following:
- generating only these effects that are produced by agent's effectors,
- delivering only these stimuli that may be observed by agents' sensors,
- size of world limited by range of variables in Java, and practically by the size of memory accessible for the simulator;

**Objects**

Every object (agent, element of environment, resource) is described by a number of attributes. Among the most important are surface coordinates. Agent algorithms may either make use of the knowledge of agent's position or not. Other methods than absolute coordinates may be used to determine position (for instance these based on *odometry*, or counting turns of wheels).

Elements of a map and resources are objects of Environment Map application. They reside on one Java virtual machine. Agents reside on the other virtual machines (to provide a possible distribution of the platform). Therefore, an agent is not really the part of the Environment Map application. Agent is represented in Environment Map application by its peer. A peer is brought to existence when an agent registers itself in the simulator (Environment Map). Further communication between server (simulator) and agent is made through its peer (Fig. 3).

**Agent**

The structure of agent is designed to reflect features of physical agent. Agent experiences other object and influences on them only through devices. Virtual agent may be substituted by physical one if real devices correspond to simulated ones.

A base of implementation of agent is the class Base Agent. It allows to initiate the agent by:
- connection to simulator (and disconnection if needed),
- definition of initial position,
- adding and removing devices,
- definition of attributes (e.g. color),

Connection and disconnection are the only interactions with simulator performed beyond devices. During the operation of the agent, Base Agent is responsible for preparation of a log and for displaying information on agent's state on the screen (if configured to do so).

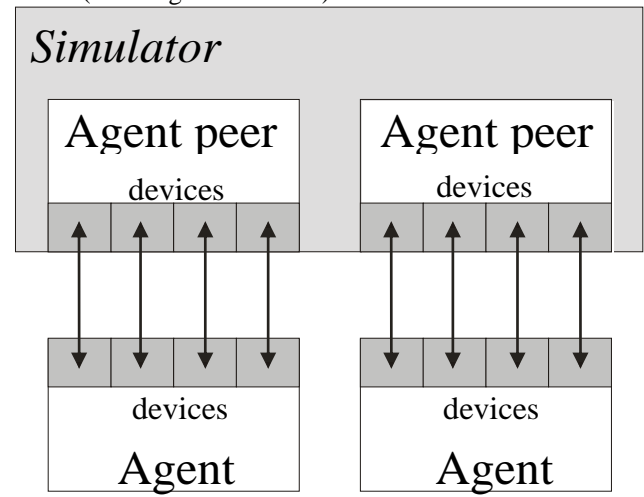

Figure 3: Agents in JADE Simulator

**EXPERIMENTS**

First, several simulations were performed with scenarios following a simple children's games. During these simulations, the JADE platform proved to be efficient, reliable and simple to use piece of software. In most cases CSM automata were used to define algorithms of individual agents.

Fig. 4 shows a simulation in which a *chasing game* is played. A trace of activities of an agent is shown. An automaton defining an algorithm of agent's behavior is presented in Fig. 5. In the simulation, agents report to the map their absolute positions. Map informs agents about directions to other agents. If an agent is very close to another one, it is informed about the distance between them. Agent is equipped with a 'touch sensor' to be informed that it has 'catched' another agent. The inertia of agents is also simulated: every order to change agent's velocity is executed with some delay.

In the simulation of a *maze* (Fig. 6), a single agent is equipped with (simulated) front microswitch sensor and two whiskers with infrared sensors allowing the robot to move in parallel to an obstacle.

Simulations of other scenario ('picking mushrooms') gave some teaching result, which made us doubt about agents' moral values. In this case, each of four agents had its 'home' located at the corner of the map. The whole map was the 'forest', where small passive objects ('mushrooms') were distributed . Each agent's task was to walk randomly around the forest and whenever a mushroom was seen within agent's range of visibility - to pick it up and bring it back home. For a new tour, at least two strategies were possible: either to continue purely random seek or to get first to the place where the last mushroom was found. Of course, other agents did the same so that the result of the quest was always uncertain. To avoid collisions, agents were not allowed to get closer to each other than some predefined distance.

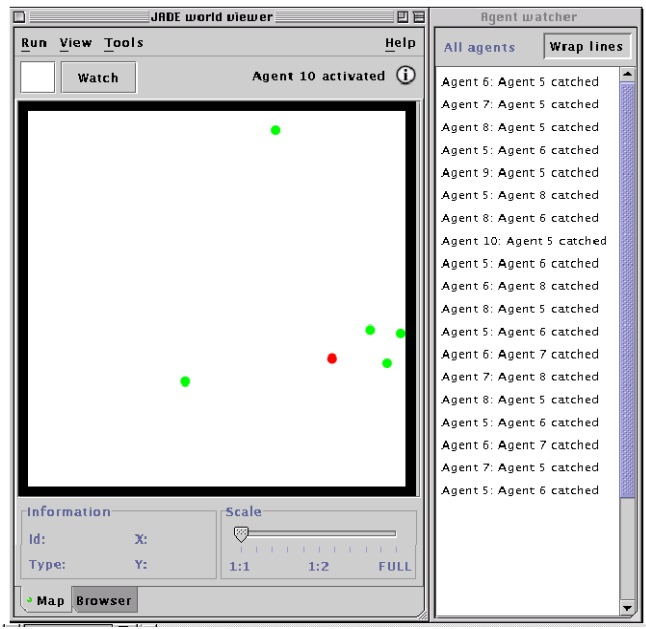

Figure 4: Chasing Game

This set of behavioral rules was easily implemented on the JADE platform and a number of simulations have been done. Agents performed as expected: they nicely collected their crop, politely avoiding mutual collisions. Each run was terminated whenever the last mushroom was removed from the forest. However, during a presentation of the platform in a seminar someone has asked: 'what would they do if the run continues while there are no more mushrooms in the forest?'. The result was predictable but still surprising and teaching: sooner or later agents become thieves, stealing mushrooms from their neighbors' homes.

This behavior can be easily explained, of course. Mushrooms brought home were made invisible to the host of this particular home, otherwise the host would never go out to seek for new ones. However, they still remained visible to other agents. Incidentally, during the first simulations it has never happened that an agent went to its neighbor's home closely enough to discover the valuable goods stored inside. Now – as simulation time was much longer – it was practically inevitable. Moreover, an agent preferring the strategy of return to the place where the mushroom was found yet – soon would become the 'professional thief', who chooses to steal permanently goods from wealthy neighbor's store instead of seeking them in the forest, even if they are still out there. Note that due to the 'anti-collision' mechanism thieves never commit their crime if the host is at home or nearby: they wait patiently or wander around until nobody is home. Also, hosts seem to be afraid of thieves: returning home they hesitate and wait until the burglar is away. It was real fun to observe this busy mini-society of decent mushroom pickers and clever thieves, all resulting from the set of simple behavioral rules implemented on the JADE platform.

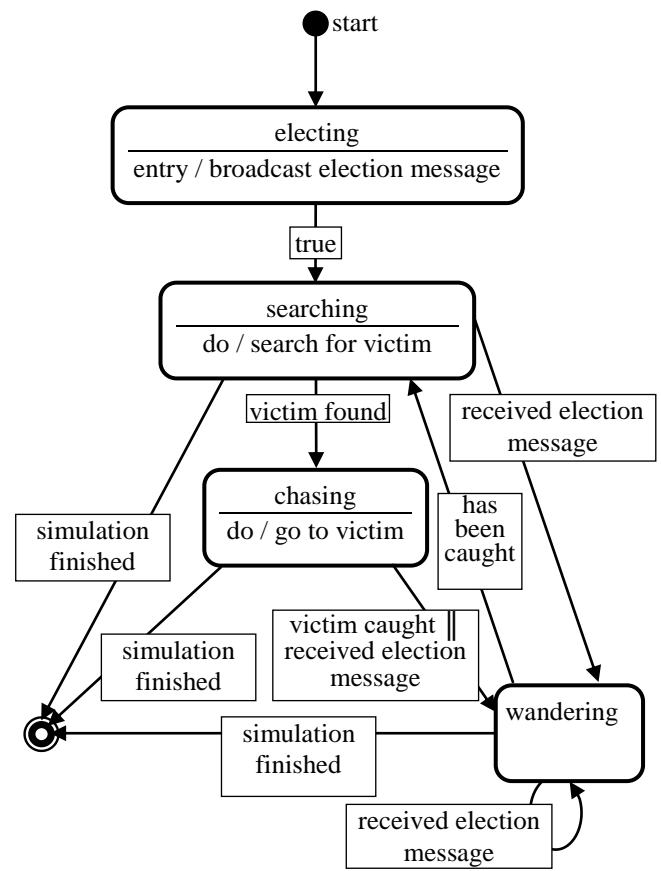

Figure 5: The Algorithm of an Agent Playing Chasing Game

Presently, the more challenging scenarios are investigated, involving the cooperation among agents toward the common goal instead of purely 'selfish' behavior. By now, two simple team tasks have been simulated:
- forming a circle (or, more precisely, a regular polygon) from a number of agents,
- building a chain (or 'a bridge over the river') from agents.

In both cases, agents knew directions and distances to other agents, but not necessarily to *all* of them. Even in the case of such simple tasks, if there is no leader (appointed or elected) and any agent makes its individual decisions as to its own movements concurrently to (and independently of) others – the solution is far from simple. The main issue is to provide the convergence of the global performance of the group toward the common goal. 'Common sense' approach yields rather poor results and more elaborate analysis of distributed algorithms is necessary.

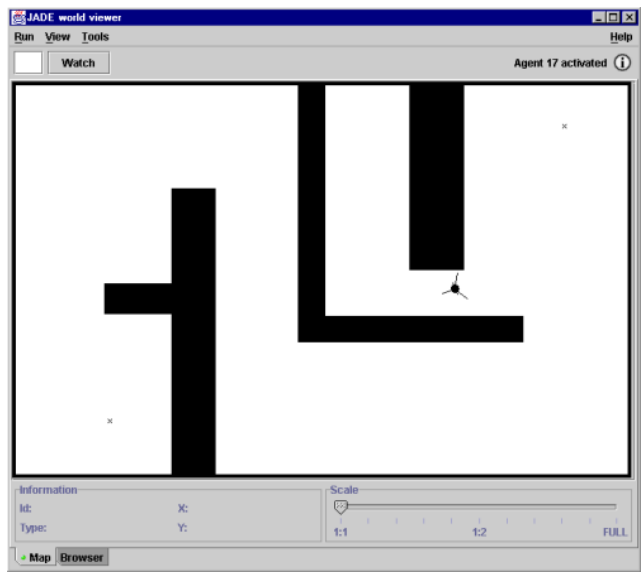

Figure 6: A Maze Simulation

## CONCLUSIONS

JADE has proved to be an efficient multiagent platform for cooperation of physical and virtual agents. It provides the conceptual and technological (software) framework for the definition of the multiagent system as well as the properties of their 'world'. However, the implementation of JADE was only a necessary prerequisite for the 'actual' research on the social behavior of groups of agents. Now, the emphasis will be put on the following issues:
- Design of new, illustrative scenarios of agents' social behavior, requiring for the use of operation and strategy layers of an agent,
- Convergence analysis of distributed algorithms (involving also the game- and decision- theoretical issues),
- Formal specification of agent's behavior (using a nested CSM model),
- Methodology of experiments (experiment planning, performance criteria, construction of sophisticated behavioral models from reusable 'building blocks' etc.)

Also, we hope that a new family of physical mini-robots cooperating with JADE platform will be ready soon. This would allow us to perform the full-scale experiments on the cooperation among virtual and real agents.

## ACKNOWLEDGMENTS

JADE platform was a subject of M.Sc. thesis of Mr. Norbert Pabiś and Mr. Paweł Zabrowski in ICS WUT. Wiktor B. Daszczuk have supervised the thesis.

This paper was partially supported by the grant No. 502/G/1032/2000/000 from the Dean of Faculty of Electronics and Information Technology, Warsaw University of Technology